# Instability Prediction in Smart Cyber-physical Grids Using Feedforward Neural Networks

Amirreza Jafari
Dept. of Elec. & Comp. Eng.
University of Tabriz
Tabriz, Iran
ar_jafari95@ms.tabrizu.ac.ir

Farzad Darbandi
Dept. of Elec. & Comp. Eng.
University of Tabriz
Tabriz, Iran
f.darbandi95@ms.tabrizu.ac.ir

Hadis Karimipour[†]
School of Engineering
University of Guelph
Guelph, Canada
hkarimi@uoguelph.ca

*Abstract*— Due to the use of huge number of sensors and the increasing use of communication networks, cyber-physical systems (CPS) are becoming vulnerable to cyber-attacks. The ever-increasing complexity of CPS bring up the need for data-driven machine learning applications to fill in the need of model creation to describe the system behavior. In this paper, a novel stability condition predictor based on cascaded feedforward neural network is proposed. The proposed method aims to identify anomaly due to cyber or physical disturbances as an early sign of instability. The proposed neural network utilizes cascaded connections in order to increase accuracy of the prediction. The conjugate gradient backpropagation and Polak-Ribière formula are utilized for training process. This method also can predict the critical generators to mitigate the effect of the cascading failure and consequent blackout in the system. Simulations results on the IEEE 39-bus system indicate the superiority of the proposed method in terms of accuracy, speed, and robustness.

*Index Terms*--cascaded feedforward neural network, deep learning, transient stability, power system, protection.

## I. Introduction

With the integration of smart metering infrastructure and communication network, traditional power systems are transformed into the smart cyber-physical systems (CPS). The emerging smart power grids are among the most critical CPSs that revolutionized energy production and management. The ever-increasing complexity of CPS bring up the need for data-driven machine learning applications to fill in the need of model creation to describe the system behavior [1]. Because in many CPSs including smart power grids reliable performance of the system is a vital requirement, it is important to detect anomalous or faulty system behavior in real time.

In the normal operation, the synchronous generators operate in harmony with together [2]. Various cyber-attacks can manipulate the system in a way that generators lose their synchronism that If not properly addressed, out of step (OOS) condition occurs which results in instability and severe damages in the system [3]. Therefore, the early prediction of the cyber or physical disturbances and the OOS condition is essential to prevent these problems.

Diverse cyber-attack detection and prevention methods have been proposed in the literature that utilizes deep learning and machine learning [4-5]. the combined feature selection algorithm [6], robust massively parallel dynamic state estimation [7]

Various OOS prediction methods have been presented in references with different advantages and defects. These methods assess the current or future transient stability (TS) condition of the smart power systems based on the various parameters behavior. Transient stability assessment (TSA) methods can be categorized into two parts, as follows:

1) The methods of first group detects transient after OOS condition. the utilization of the distance relays [8], phasor-based out-of-step detectors and dual blinder [9] and the wavelet transform [10] are the most common techniques of this group. The possibility of the incorrect identification between fault and the OOS condition and high-latency in detection are the main disadvantage of these methods.

2) The second groups include methods that predicts the instability of the power system at future moments based on the current behavior of the system parameters. These methods commonly utilize intelligent and deep learning based techniques. Since these methods predict the stability condition of the power system before OOS, are more efficient and applicable than first group methods.

The methods of this group include support vector machine [11], Bayesian technique [12], fuzzy theory [13], decision tree (DT) [14], state estimation [15], extreme learning machine [16] and ensemble online sequential learning machine [17]. Among these methods, artificial neural network (ANN) based methods are assumed as the most reliable, fast and integrated approaches, which can handle the complicated pattern recognition problems of the power system with big data. The convolutional neural network (CNN) [18] and cascaded convolutional neural network (CCNN) [19] are the examples of deep learning based methods that have been presented for TSA.

Majority of the mentioned methods consider one type of fault and from this point of view are not practical in the real systems. In addition, due to weak structure of used neural network, these methods have low accuracy and have to utilize post fault data in order to increase accuracy up to acceptable level that lead to the late prediction.

In this paper, a new structure of feedforward neural network with cascaded connections is designed for TSA in the smart power systems. In order to design the cascaded feedforward neural network (CFNN), the number of cascaded connections is selected optimally that improves the accuracy of the network. The conjugate gradient backpropagation (CGB) and Polak-Ribière formula have been used for learning process. The sampling strategy for input vector generation has been conducted in such a way that, whole fault duration and some moments of system initial state are considered in input vectors. The presented method considers TSA of all fault types therefore can be considered as a comprehensive method for all short circuit types. On the other hand, due to robust structure and considering cascaded connections in an optimal way, the designed CFNN has high accuracy and fast performance.

The rest of paper is organized as follows. Section 2 introduces the TS. Section 3 proposes CFNN algorithm, feature sampling strategy, and training model of network. Simulation results and sensitivity analysis are discussed in Section 4. Finally, Section 5 concluded the paper.

## II. Transient Stability Assessment

TS is the ability of system to retain synchronism after contingencies. In this condition, synchronous generators preserve their synchronism with together and their rotor angle differences remain less than 180 degrees. Short circuit faults are the most common cause for the loss of stability in the power systems. In a typical fault scenario, the faulty component is disconnected by the action of the adjacent breakers. During this time, the system moves away from the pre-fault balanced point and experiences transient fluctuations. In the unstable cases, undamped oscillations result in power swing between generators, which is named the OOS condition. In this situation, the intensity of the oscillations rapidly increases which may cause irreparable damage to the system and even results in blackout in whole power system.

To protect the system, TSA needs to be conducted to predict the instability. It is obvious that a faster prediction is always preferred to provide extra time for control actions. Therefore, fast and accurate identification of the OOS condition is paramount for the stable and reliable operation of the power systems.

### A. Stability Detection

Traditionally, TS detection is performed once OOS condition happened. The most common technique is utilizing distance relays to measure the impedance variation rate in the generators. By examining the entrance velocity of the impedance from one zone to another zone and fault duration, the instability of the power system can be detected. The main drawback of these approaches is their high latency, which usually results in severe damages to the equipment and spreading instability over the system.

### B. Stability Prediction

Another approach for TSA includes prediction-based techniques that attempt to predict instability before OOS happens. These methods predict the future stability condition by analyzing the current behavior of the system parameters. This provides enough time to take action and mitigate the effect of instability in the system. Among various techniques, machine learning-based methods have shown promising results. This paper proposes an accurate method based on deep learning for online prediction of the power system instability.

## III. CFFN-based TSA

Various neural network structures are used for evaluation and prediction of the state of the systems by analyzing their signs. Depending on the type of the problems and the systems, these structures can be transformed into the optimal forms. The feedforward neural network (FNN) is considered as a one of the most used deep learning methods. FNN is composed from input, hidden and output layers. In each layer number of units called 'neuron' have the duty of generating outputs from the inputs of the previous layer. In FNN, the connections between layers exist only between consecutive layers and the procedure of network takes place in a forward direction. One of the structures of FNN is cascaded FNN

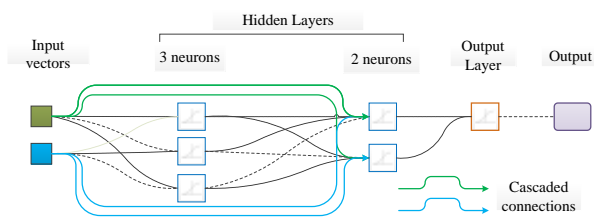

(CFNN) that its general scheme is shown in fig. 1.

Figure 1. Cascaded feedforward neural network

As seen in fig. 1, CFNN in addition to consecutive connections utilizes forward connections between input vectors and all hidden layers. This network prognosticates output classes according to input values. By using the external connections, each neuron adjusts its output value based on the previous layer outputs and main input vectors, and as a result enhance the accuracy of prediction.

### A. Proposed CFNN Algorithm

This paper uses CFNN for TSA of power system. In the designed CFNN, CGB [20] and the Polak-Ribière formula [21] are used for learning and searching process, respectively. The CGB that has high convergence speed and needs low computational memory, is performed in the sparse systems that are complicated to be solved by direct methods such as the Cholesky decomposition.

Generally, the CGB is performed iteratively. In the first iteration, the algorithm searches the solution space in the gradient descent direction. In the next iterations, the searching in the direction of the negative of the gradient does not guaranty the fastest convergence. Therefore, in the next iterations, the searching process is done on the direction combined of previous iteration direction and current steepest descent direction, as (1):

$$R_k = -G_k + P_k^f R_{k-1} \quad (1)$$

In (1), $G_k$ and $R_k$ are gradient vector and searching direction vector in the $k^{th}$ iteration, respectively. $P_k^f$ is the searching index corresponding to Polak-Ribière formula that can be obtained according to following equation:

$$P_k^f = \frac{G_k^T (G_k - G_{k-1})}{G_{k-1}^T G_{k-1}} \quad (2)$$

Where $G_k^T$ is the transposed form of the gradient vector at $k^{th}$ iteration. In the next iterations, the changing distance of the algorithm is calculated as (3):

$$x_{k+1} = x_k + \delta_k R_k \quad (3)$$

In (3), $\delta_k$ determines the step length of the algorithm at $k^{th}$ iteration that is obtained through a linear searching in the conjugate direction in order to minimizing objective function value ($f(x)$), according to:

$$\delta_k = \min_{\delta} f(x_k + \delta_k R_k) \quad (4)$$

The cross-entropy function is calculated between actual and predicted output of the network. This error is used as a termination criterion in order to assess the learning level. In addition, for developing an appropriate correlation between input and output vectors, the input vectors are normalized that improves the performance of the CFNN.

One of the main properties of the CFNN is the utilization of the cascaded connections in its structure. In some cases, these connections increase the performance of the network, but in some others increase the computational burden and has not considerable impact on the accuracy of network. For designing optimal structure of the CFNN, the presence of each cascaded connection is considered as a binary variable, and the resulted optimization problem is solved by branch and bound method. In this optimization problem, the performance function of the CFNN is assumed as the objective function.

### B. Feature Sampling

Implementation of the designed CFNN for TSA of the power system requires a set of suitable features that the network input vectors are produced by sampling of them. These features include mechanical and electrical parameters of the power system and generators. In order to achieve the efficient performance of the CFNN, the set of features should be selected optimally. In this paper, various features such as stator current, magnitude and angle of stator voltage, excitation current of generator, rotor speed, rotor angle differences, active and reactive power of generators and load angle, are separately examined and the set of features are selected among them, according to their performance.

Input vectors of the designed CFNN are generated by sampling of the selected features. For sampling these features, a time window with a specific length and sampling frequency are chosen. The length of time window is selected adequately to include whole fault duration and some moments of the system initial state. The number of samples in input vectors is calculated by (5).

$$N_S = F_S \times LTW \quad (5)$$

Where, $N_S$, $F_S$ and $LTW$ are the number of samples in the input vectors, the sampling frequency and the length of time window, respectively.

### C. Training Designed CFNN

After selection of the set of features and generating input vectors, the designed CFNN with two sub-function is trained. The first function predicts the stability condition of the power system based on the input vectors, and the second function predicts the pair of critical generators that are most sensitive to desired fault and go to the OOS before other generators. For training the first function, the actual state of the system in post-fault moments and for the second function the index of the critical generators are given to the output vector of the CFNN.

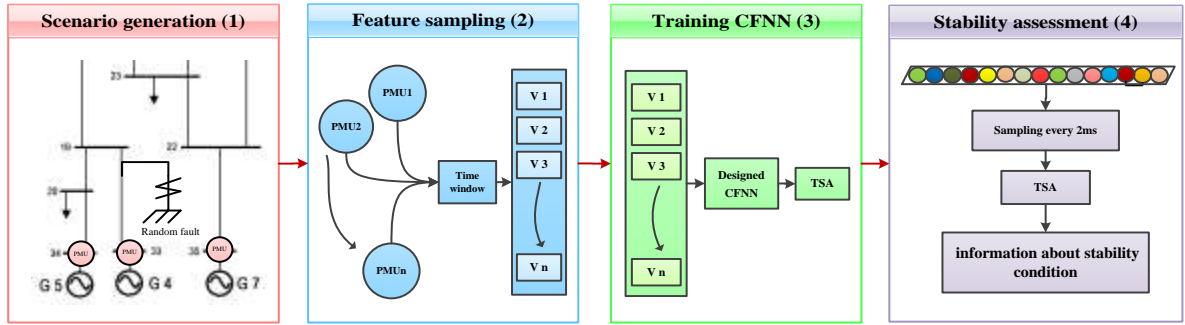

Figure 2. The flowchart of proposed TSA methodology

*D. Implementation of CFNN for TSA*

In order to perform the designed CFNN, the sampling process is carried out constantly, and the designed time window always latches the specific number of derived samples in its memory. Whenever a clearance sign is received from the adjacent breakers of occurred fault, the sampling process is halted, and the latest input vectors are given to the trained function. The running of functions takes shorter than 1ms, therefore proposed TSA process is performed at maximum 1ms plus the duration of data communication after fault clearance. Fig. 2 shows the general scheme of proposed method. As seen in fig. 2, the proposed algorithm starts by generating scenarios for all fault types. In the next step, optimal feature set, are composed from available features, and by utilizing determined time window, the selected features are sampled, and the input vectors are generated. In the step (3), the designed CFNN is trained based on generated input vectors and two sub-functions are obtained. In the last step, trained functions are implemented in the system and predict the future stability condition and critical generators. As seen in fig. 2, the proposed algorithm starts by generating scenarios for all fault types. In the next step, optimal feature set, are composed from available features, and by utilizing determined time window, the selected features are sampled, and the input vectors are generated. In the step (3), the designed CFNN is trained based on generated input vectors and two sub-functions are obtained. In the last step, trained functions are implemented in the system and predict the future stability condition and critical generators.

## IV. RESULT AND DISCUSSION

The proposed TSA algorithm are tested on IEEE new-England test system. This system has 39 buses and 10 generators. The simulations are done in the MATLAB software. The selected feature set includes output active power, magnitude of stator voltage and rotor speed of generators. By considering 10 connected generators and 3 features for each of them, the total number of features will be equal to 30 features. The number of hidden layers in the CFNN are considered equal to 20 layers. The length of time window and the sampling frequency are 500 ms and 500 sample per second, respectively. Therefore, each input vector will contain 250 samples.

In the scenarios, all fault types include 3-phase to ground, 2-phase, 2-phase to ground and single-phase to ground faults, are considered. In addition, fault duration is selected randomly between 60 and 400ms. Probable faults can be occurred along all lines of the power system and the load profile can vary between 70 to 140 percent of the basic load. In order to train and test designed CFNN, 1000 and 500 scenarios have been generated, respectively.

*A. Performance Analysis*

For assessing efficiency of the proposed TSA method, two sub-functions are trained by 1000 training scenarios and are tested on 500 test scenarios. Table I illustrates the results of first function related to stability prediction, compared to the other methods in the literature.

TABLE I.   THE ACCURACY RESULTS OF STABILITY PREDICTION FUNCTION

| Prediction algorithm | accuracy % | Prediction Time | Fault Type |
|---|---|---|---|
| Bayesian Technique [12] | 91.57 | 100 ms after fault clearance | All types |
| DT [14] | 90.3 | 50 ms after fault clearance | 3-phase |
| FC [22] | 97.15 | 70 ms after fault clearance | All types |
| CNN [18] | 89.22 | 90 ms after fault clearance | 3-phase |
| Proposed TSA method | 97.8 | 1 ms after fault clearance | All types |

As can be seen from Table I, the proposed TSA method predicts stability condition at one ms after fault clearance. In addition, the average accuracy of this function is 97.8% that has superiority to other methods, which predict the OOS condition at a longer time after fault clearance with lower accuracy. The lower prediction time of presented method is the significant factor that can help protection strategies to save stability of the system more efficiently. It should be mentioned that in majority of the existing techniques only one types of the fault is analyzed while proposed method significantly outperform existing techniques in terms of accuracy, speed and integrity. Fig. 3 shows a random 3-phase fault and the real-time output of stability predictor function for this fault.

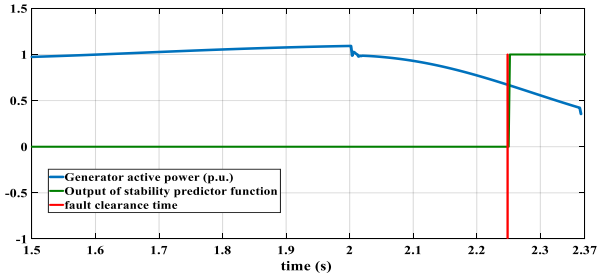

Figure 3.   Output of stability predictor function for 3-phase fault

As seen in Fig. 3, the fault occurred at t=2s and has been cleared at t=2.249s. The stability predictor function acts at 2.25s (one millisecond after fault clearance). To predict the pair of critical generators, which will result in OOS condition before other generators, another sub-function is trained. This function has been trained by 1000 training scenarios and has been tested on 500 additional scenarios. The training and testing accuracy of this function has been calculated 98.2 and 97.5 %, respectively.

This function localizes the effect of fault by isolating critical generator and saves lots of time for fault mitigation through proper protection schemes. Fig. 4 shows the output curve of this function for a random 2-phase to ground fault. As seen in fig. 4, critical generator estimator function predicts second and eighth generators as critical generators at 1ms after fault clearance time.

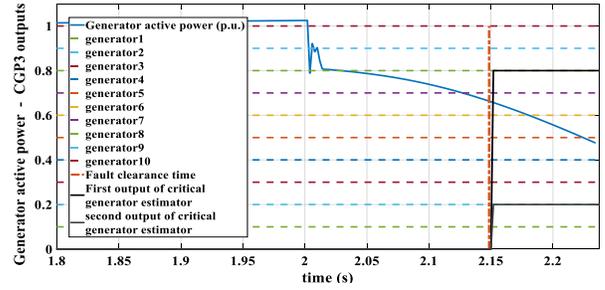

Figure 4.   Output variation of critical generator predictor function for a random 2-phase fault.

The rotor angle difference of the generators is also presented in Fig. 5. As can be seen, the rotor angle difference of second and eighth generators has reached to 180 degrees faster than other pairs of the generators.

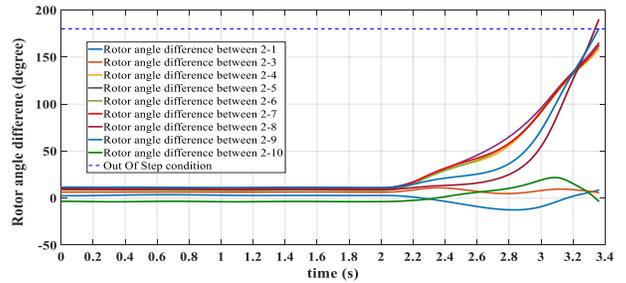

Fig. 5 Rotor angle difference of generators during 2-phase fault

### B. Sensitivity Analysis

The presence of cascaded layers can enhance the performance of the designed network. These connections are used between input vectors and hidden layers, therefore the maximum number of them is equal to multiple of sum of neuron units in the hidden layers except first hidden layer (48) by number of input vectors (30). In this paper, the total number of neurons in the designed CFNN and first layer are 51 and 3 units, respectively. In this section, a sensitivity analysis is done on the performance of the designed CFNN, based on the number of cascaded connections.

TABLE II.   THE ACCURACY AND FPR OF PROPOSED TSA METHOD IN DIFFERENT NUMBER OF CASCADED CONNECTIONS

| Num of cascaded connections | 90 | 180 | 270 | 360 | 450 | 540 | 630 | 720 |
|---|---|---|---|---|---|---|---|---|
| Accuracy (%) | 95.6 | 95.4 | 96.8 | 97.3 | 97.3 | 97.4 | 97.8 | 97.5 |
| FPR | 0.063 | 0.067 | 0.048 | 0.032 | 0.031 | 0.028 | 0.019 | 0.026 |
| Num of cascaded connections | 810 | 900 | 990 | 1080 | 1170 | 1260 | 1350 | 1440 |
| Accuracy (%) | 97.6 | 97.3 | 97.1 | 96.4 | 96.7 | 96.2 | 96.0 | 96.1 |
| FPR | 0.024 | 0.031 | 0.036 | 0.053 | 0.049 | 0.056 | 0.059 | 0.057 |

Table II shows the accuracy and false positive rate (FPR) of the CFNN in the different number of cascaded connections. As seen in the table, by increasing the number of cascaded connections up to 630, the accuracy of proposed method has been increased to 97.8%, but by increasing the number of connections higher than 630, the accuracy of TSA method has been gradually decreased. In addition, the results show that the minimum FPR has been achieved when 630 connections are considered in the CFNN.

## V. Conclusion

In this paper, a novel CFNN based TSA method has been presented that can predict instability of the smart power systems with high accuracy and fast response. The designed CFNN, is trained by pre-fault samples include whole fault duration and some moments of system initial state and is independent of post-fault data. The running time of the presented method is lower than 1 ms, therefore the proposed method can predict future stability condition in one ms after fault clearance. In the presented CFNN, in order to increase performance of assessment, the number of cascaded connections is selected optimally and this process enhance the robustness of the designed structure. The CGB and Polak-Ribière updates method are used to learning step. In addition, the proposed method has the capability of predicting critical generators, which helps protection methods to select appropriate strategies for prevention of instability in the system. The simulation have been done on IEEE 39-bus system, and the obtained results indicates that the proposed method with 97.8 percent accuracy and lower prediction time are superior to other methods.

## References

[1] J. Sakhnini, H. Karimipour, and A. Dehghantanha, "Smart grid cyber attacks detection using supervised learning and heuristic feature selection," *IEEE SEGE, pp. 1-5*, Aug. 2019.

[2] H. Karimipour, V. Dinavahi, "Parallel Domain Decomposition Based Distributed State Estimation for Large-scale Power Systems", IEEE Transactions on Industry Applications, vol. 52, no. 2, pp. 1265-1269, March 2016.

[3] H. Karimipour, V. Dinavahi, "Extended Kalman Filter Based Massively Parallel Dynamic State Estimation", IEEE Transaction in Smart Grid, vol. 6, no. 3, pp.1539-1549, May 2015.

[4] H. Karimipour, A. Dehghantanha, R. M. Parizi, K. R. Choo, and H. Leung, "A deep and scalable unsupervised machine learning system for cyber-attack detection in large-scale smart grids," *IEEE Access, vol. 7, pp. 80778-80788*, 2019.

[5] A.N. Jahromi, S. Hashemi, A. Dehghantanha, K.-K.R. Choo, H. Karimipour, D.E. Newton, R.M. Parizi, "An improved two-hidden-layer extreme learning machine for malware hunting", Comput. Secur., vol. 89, pp.1-11 ,2019.

[6] S. Mohammadi, H. Mirvaziri, M. Ghazizadeh-Ahsaee, and H. Karimipour, "Cyber intrusion detection by combined feature selection algorithm," *Journal of information security and applications*, vol. 44, pp. 80–88, 2019.

[7] H. Karimipour and V. Dinavahi, "Robust massively parallel dynamic state estimation of power systems against cyber-attack," *IEEE Access*, vol. 6, pp. 2984–2995, 2017.

[8] U. J. Patel, N. G. Chothani, and P. J. Bhatt, "Distance relaying with power swing detection based on voltage and reactive power sensitivity," *International Journal of Emerging Electric Power Systems*, vol. 17, no. 1, pp. 27–38, 2016.

[9] M. A. Saad, A. H. Eltom, G. L. Kobet, and R. Ahmed, "Performance comparison between Dual-Blinder and Phasor-Based Out-of-Step detection functions using hardware-in-the loop simulation," in *2015 IEEE Industry Applications Society Annual Meeting*, 2015, pp. 1–8.

[10] M. R. Salimian, F. Haghjoo, and M. R. Aghamohammadi, "Out of step detection and protection using online WT," *International Electrical Engineering Journal (IEEJ)*, vol. 4, no. 4, pp. 1133–1139, 2013.

[11] L. S. Moulin, A. P. A. Da Silva, M. A. El-Sharkawi, and R. J. Marks, "Support vector machines for transient stability analysis of large-scale power systems," *IEEE Transactions on Power Systems*, vol. 19, no. 2, pp. 818–825, 2004.

[12] H. Zare, Y. Alinejad-Beromi, and H. Yaghobi, "Intelligent prediction of out-of-step condition on synchronous generators because of transient instability crisis," *International Transactions on Electrical Energy Systems*, vol. 29, no. 1, p. e2686, 2019.

[13] I. Kamwa, S. R. Samantaray, and G. Joós, "On the accuracy versus transparency trade-off of data-mining models for fast-response PMU-based catastrophe predictors," *IEEE Transactions on Smart Grid*, vol. 3, no. 1, pp. 152–161, 2011.

[14] M. R. Aghamohammadi and M. Abedi, "DT based intelligent predictor for out of step condition of generator by using PMU data," *International Journal of Electrical Power & Energy Systems*, vol. 99, pp. 95–106, 2018.

[15] Y. Cui, R. G. Kavasseri, and S. M. Brahma, "Dynamic state estimation assisted out-of-step detection for generators using angular difference," *IEEE Transactions on Power Delivery*, vol. 32, no. 3, pp. 1441–1449, 2017.

[16] R. Zhang, Y. Xu, Z. Y. Dong, and K. P. Wong, "Post-disturbance transient stability assessment of power systems by a self-adaptive intelligent system," *IET Generation, Transmission & Distribution*, vol. 9, no. 3, pp. 296–305, 2015.

[17] H. Yang, W. Zhang, F. Shi, J. Xie, and W. Ju, "PMU-based model-free method for transient instability prediction and emergency generator-shedding control," *International Journal of Electrical Power & Energy Systems*, vol. 105, pp. 381–393, 2019.

[18] A. Gupta, G. Gurrala, and P. S. Sastry, "An online power system stability monitoring system using convolutional neural networks," *IEEE Transactions on Power Systems*, vol. 34, no. 2, pp. 864–872, 2019.

[19] R. Yan, G. Geng, Q. Jiang, and Y. Li, "Fast transient stability batch assessment using cascaded convolutional neural networks," *IEEE Transactions on Power Systems*, 2019.

[20] L. M. Saini and M. K. Soni, "Artificial neural network-based peak load forecasting using conjugate gradient methods," *IEEE Transactions on Power Systems*, vol. 17, no. 3, pp. 907–912, 2002.

[21] L. Zhang, W. Zhou, and D.-H. Li, "A descent modified Polak–Ribière–Polyak conjugate gradient method and its global convergence," *IMA Journal of Numerical Analysis*, vol. 26, no. 4, pp. 629–640, 2006.

[22] M. Chen et al. "XGBoost-based algorithm interpretation and application on post-fault transient stability status prediction of power system," *IEEE Access*, vol. 7, pp. 13149–13158, 2019.